\documentclass[runningheads]{llncs}
\usepackage{graphicx}
\usepackage{listings}
\usepackage[numbers]{natbib}
\usepackage{hyperref}
\usepackage{todonotes}
\usepackage{array}
\newcolumntype{P}[1]{>{\centering\arraybackslash}p{#1}}
\begin{document}
\title{EduCOR: An Educational and Career-Oriented Recommendation Ontology}
%
%
\author{Eleni Ilkou\inst{1}\orcidID{0000-0002-4847-6177} 
\and
Hasan Abu-Rasheed\inst{3}\orcidID{0000-0002-2921-4809} 
\and
Mohammadreza Tavakoli\inst{1,2}\orcidID{0000-0002-7368-0794}
\and
Sherzod Hakimov\inst{2}\orcidID{0000-0002-7421-6213}
\and
G\'abor Kismih\'ok\inst{2}\orcidID{0000-0003-3758-5455}
\and
Sören Auer\inst{2}\orcidID{0000-0002-0698-2864}
\and
Wolfgang Nejdl\inst{1}\orcidID{0000-0003-3374-2193}
}
\authorrunning{E. Ilkou et al.}
%
\institute{L3S Research Center, Leibniz University Hannover, Germany
\\
\and
TIB -- Leibniz Information Centre for Science and Technology, Hannover, Germany
\and WBS\&WM Institute, University of Siegen, Siegen, Germany\\
\email{\{ilkou, tavakoli, nejdl\}@l3s.de} \\
\email{\{sherzod.hakimov, gabor.kismihok, auer\}@tib.eu}\\
\email{\{hasan.abu.rasheed\}@uni-siegen.de} \\
}
\maketitle              
\begin{abstract}
With the increased dependence on online learning platforms and educational resource repositories, a unified representation of digital learning resources becomes essential to support a dynamic and multi-source learning experience. We introduce the EduCOR ontology, an educational, career-oriented ontology that provides a foundation for representing online learning resources for personalised learning systems. The ontology is designed to enable learning material repositories to offer learning path recommendations, which correspond to the user's learning goals, academic and psychological parameters, and the labour-market skills. 
We present the multiple patterns that compose the EduCOR ontology, highlighting its cross-domain applicability and integrability with other ontologies. A demonstration of the proposed ontology on the real-life learning platform eDoer is discussed as a use-case. We evaluate the EduCOR ontology using both gold standard and task-based approaches. The comparison of EduCOR to three gold schemata, and its application in two use-cases, shows its coverage and adaptability to multiple OER repositories, which allows generating user-centric and labour-market oriented recommendations.

\textbf{Resource}: \url{https://tibonto.github.io/educor/}



\end{abstract}

\section{Introduction} \label{intro}


In recent years, digital education is increasingly relying on Educational Resources (ERs) and Open Educational Resources (OER). These ERs are available in many different formats, such as videos, slide decks, audio recordings from lectures, digital textbooks, or simple web pages. Furthermore, ERs and OERs usually come with low-quality metadata~\cite{tavakoli2021metadata}, and they are isolated from other, content-wise similar ERs. That is one of the important reasons for lacking high-quality services (e.g. recommendation and search services) based on OERs~\cite{tavakoli2020quality}. Therefore, it is not surprising that the Semantic Web (SW) community shows increased interest in organising and classifying ERs, and enhancing the metadata in publicly available ER and OER~\cite{duran2021integration, koutsomitropoulos2020semantic}. Although many schemata and vocabularies were suggested in the past for the educational domain, only a few of them are still available online and can accommodate particularities of OERs, and related personalised recommendation systems' features. Furthermore, recent works revealed the increased interest in educational Knowledge Graphs~\cite{dang2019mooc,ilkou2020technology}, which, however, often lack an underlying ontology or schema~\cite{chen2018knowedu}. Commercial products seem to follow a similar direction, as they often do not use or do not publish their underlying knowledge schema\footnote{An example is the Mathspace \url{https://mathspace.co}, a math education platform that offers personalised learning based on a Knowledge Graph. However, its knowledge schema is not publicly available.}. Additionally, surveys in e-learning have shown that an ontology is helpful in achieving personalised recommendation systems~\cite{GeorgeL19, TarusNM18}. Moreover, there is an increased interest on the education side to enrich current tools with Artificial Intelligence to achieve Smart Education. In this line, ontologies offer a wide variety of benefits for Smart Tutoring Systems~\cite{SalemN19}. In addition, the SW 
has a significant focus on question answering and (learning) recommendation systems. The latter is evolving rapidly to offer interoperability, explainability, and user privacy 
while providing personalised learning recommendations~\cite{barria2019explaining, chicaiza2017recommendation}. 

On the broader community side, there is strong evidence of everyday usage of online learning. Societies put enormous effort into the digital transformation of education, such as the Digital Educational Plan of the European Union\footnote{\url{https://ec.europa.eu/education/education-in-the-eu/digital-education-action-plan_en}}, on matching work and relevant skills, and on executing skill development in online learning platforms~\cite{davies2020developing}. These online learning platforms are used daily by millions of learners, especially during the COVID-19 pandemic, when education has been pushed towards online environments worldwide. Consequently, a need for lifelong learning tools emerged that could assist people in career changes, (re)skilling, or (re)entering the labour market after a period of unemployment. This trend is visible in the last decade through an increased public interest in online learning supportive platforms, such Coursera~\cite{Coursera} for lifelong learning, or Khan Academy\footnote{\url{https://www.khanacademy.org}} for school education. These platforms usually contain ERs in video format, and assessments to validate learners' knowledge, yet they also indicate new challenges by shifting learning towards personalised recommendations. 

However, this personalisation agenda of education requires novel ways to model learning processes, especially in complex learning environments. This is especially challenging when the ingredients of the learning process are originated from the angles of education (learning content and instruction), the labour market (learning context), and individual needs of learners (learning objectives). Ontologies engineered by the SW community can play a crucial role here. While there are plenty of works available, both as e-learning and occupational ontologies, no model is available currently to connect these two domains.
 
Therefore, following both SW and broader community interest, we developed the Education and Career-Oriented Recommendation Ontology, the EduCOR ontology. This syntactic formalism describes ERs, skills, and the user profile in rich metadata. It creates the bridge between the demanding and constantly changing needs of the labour market and the educational domain. EduCOR provides both the basis of an educational Knowledge Graph, and serves as a potential framework for personalised, OER recommendation systems. 
To the best of our knowledge, the EduCOR ontology is breaking new ground on modelling ERs for a personalised recommendation system based on the learner's learning path and user profile. Moreover, EduCOR is filling an essential gap in connecting personalised learning recommendation systems, educational data and skills with the labour market, making it a vital schema for future applications.

\section{EduCOR Ontology} \label{coreonto}

The EduCOR ontology is proposed to organise different domain ERs and OERs under a common ontology, link to the labour market, and offer personalised recommendation systems in the e-learning domain. A general cross-domain educational ontology should serve different purposes. Given this multidisciplinary interest and diversity of applications, there is a need for semantic representation under a unified framework that can accommodate associations between entities and attributes. 
We performed a requirement analysis for e-learning platforms to host personalised recommendations by reviewing the literature and an existing e-learning system. As a result, we identified the key components around which we constructed our ontology.

\subsection{Ontology Composition}

Our ontology introduces the necessary classes and properties to construct an e-learning environment that supports personalised recommendations. Before developing our ontology, we examined state-of-the-art related works, open standards, and best practices. 

Since our goal was to create a general ontology, we limited our conceptual work to high-level, fundamental constructs. Consequently, we examined a series of open standards related to educational content, and we critically choose those that offer a wide coverage over the narrower focused ones. Thus, we adopted the widely used IEEE LOM Standard\footnote{\url{https://standards.ieee.org/standard/1484_12_1-2020.html}} and LRMI Standard\footnote{\url{https://www.dublincore.org/specifications/dublin-core/dces}}. Furthermore, we reuse parts from the Curriculum Course Syllabus Ontology (CCSO)~\cite{KatisKAV18} and schema.org\footnote{\url{https://schema.org/}}. Furthermore, our ontology is aligned with FAIR principles~\cite{wilkinson2016fair}. Our data are assigned globally unique and persistent identifiers, and they are described with rich metadata, which is accessible and retrievable as it is demonstrated in the ontology page\footnote{\url{http://ontology.tib.eu/educor}}. We use OWL for the ontology representation, and we reuse vocabularies that follow FAIR principles and include references to them. We describe the scope of our data and have them published under the licence CC0 1.0 Universell (CC0 1.0)
Public Domain Dedication\footnote{\url{https://creativecommons.org/publicdomain/zero/1.0/deed.de}}, and it has the canonical citation ``E. Ilkou et al: EduCOR: An Educational and Career-Oriented Recommendation Ontology''.

\begin{figure}
\centering
\includegraphics[width=\textwidth]{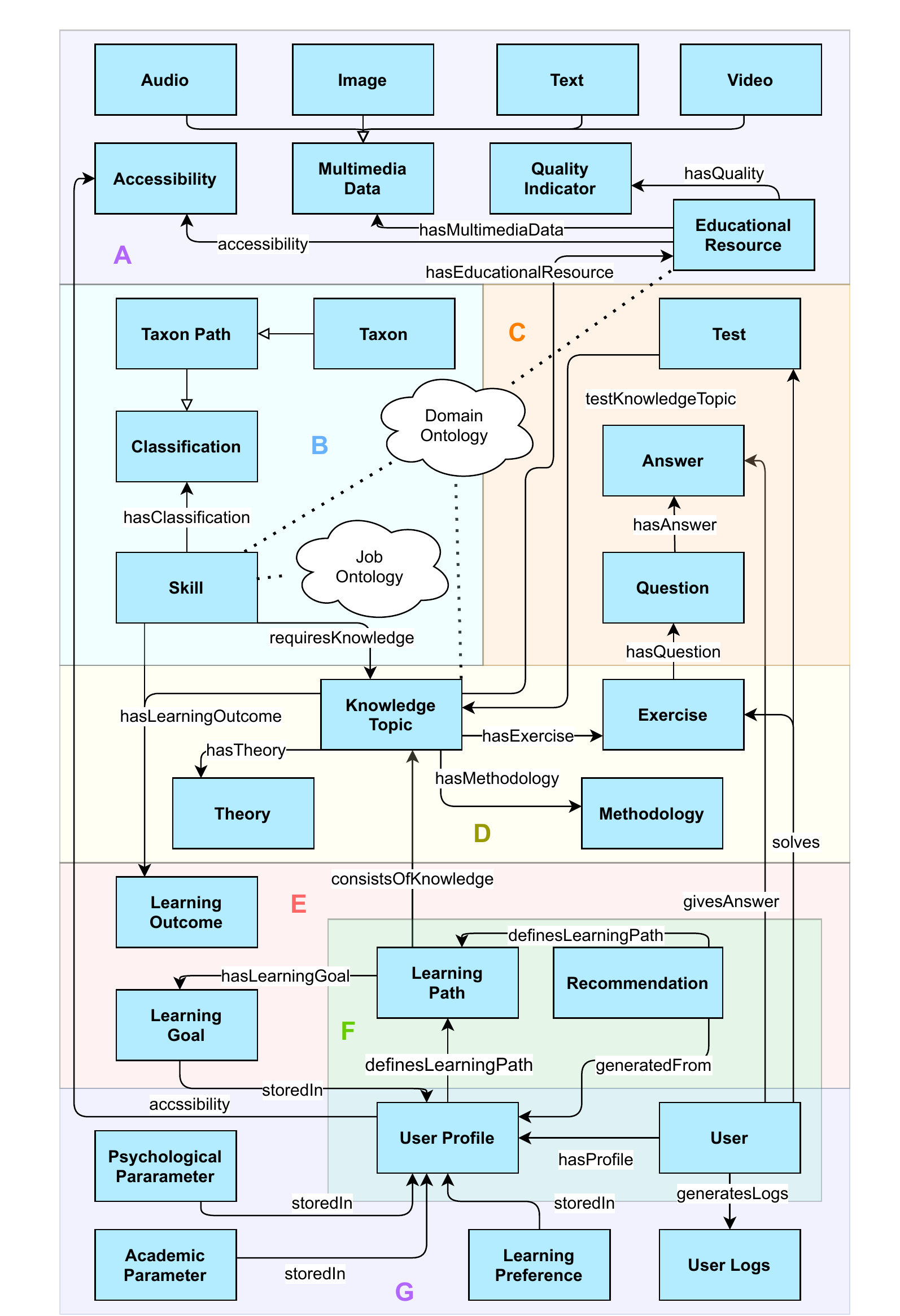}
\caption{An overview of the EduCOR ontology classes. Each pattern of the ontology is highlighted individually, \textbf{A}: Educational resource pattern, \textbf{B}: Skill pattern, \textbf{C}: Knowledge topic pattern, \textbf{D}: Test pattern, \textbf{E}: Learning path pattern, \textbf{F}: Recommendation Pattern, \textbf{G}: User profile pattern. 
} 
\label{fig1}
\end{figure}

Before finalising our design, we had an expert evaluation phase, where we received feedback from domain and ontology experts. 
The ontology also offers classes as plug-in points, where other ontologies can be mapped for more specific utilisation. Such an example is the `Learning Preference' that could host a thorough analysis as it is presented by \citet{CiloglugilI18}.
In Figure~\ref{fig1}, we present a conceptual overview of the classes in EduCOR ontology with connections to a domain ontology and job ontology. 
A comprehensive presentation of each class's object and data properties can be found on the ontology page. 

\subsection{Patterns}\label{patterns}

EduCOR consists of independent modules that can be combined to create the complete schema of the ontology. We also refer to the modules as patterns. Based on our requirement analysis, we identified the key components of a personalised learning recommendation system. Taking these components as the central theme of each module presentation, we created the additional patterns respectively. The patterns EduCOR identifies are the following: \textit{Educational Resource}, \textit{Knowledge Topic}, \textit{Skill}, \textit{Learning Path}, \textit{Test}, \textit{Recommendation}, \textit{User Profile}. 
Each pattern stands alone and can be added to another ontology, used as a single pattern separated from the 
EduCOR ontology, if an application does or does not need it accordingly. In Figure~\ref{fig1}, the classes of each pattern are represented in different colours. 

In the \textit{Educational Resource } pattern in Figure~\ref{fig1} pattern (A), the `Educational Resource' class represents the learning material or learning object. It can have multiple types that are covered by the `Multimedia Data' class. The `Education Resource' also has a `Quality Indicator', reflecting any quality measure required by the hosting content repository. Learners' different access requirements are covered through the `Accessibility' class, which represents the access rights and methods of the learning material. 

Each `Educational Resource' refers to a specific `Knowledge Topic' in \textit{Knowledge Topic} pattern (D). 
Knowledge Topics represent specific themes in a particular domain of knowledge, such as the ``Quadratic Equations" in the ``Mathematics" domain. A `Knowledge Topic' has a `Theory' and an `Exercise' content, which the learner experiences through a specific `Methodology'. The `Exercise' class is connected to both the `Knowledge Topic' and `Test' patterns.

In \textit{Test} pattern (C), the `Test' class represents the learning assessment procedure. It is composed of one or more `Exercises', which in turn have questions and corresponding answers. A `Test' can be composed of exercises that belong to many knowledge topics, skills, and domains.

Knowledge Topics are the requirements of achieving a target `Skill'. The `Skill' class, in \textit{Skill} pattern (B), is the link between knowledge topics and the labour market job ontology. 

Mastering a targeted `Skill' and `Knowledge Topic' can happen through their unique `Learning Outcome'. Such `Learning Outcome' results from the recommended `Learning Path', in \textit{Learning Path} pattern (E). The `Learning Path' pattern represents the sequence of knowledge topics needed to reach a user-defined `Learning Goal' through the intermediate `Learning Outcomes' of each `Knowledge Topic' in the recommended `Learning Path'. 

The `Recommendation' class, in \textit{Recommendation} pattern (F), is designed to cover a range of recommended item-types based on the use-case requirements. A `Recommendation' is directly generated from the `User Profile', in pattern (G), which is the means of modelling the `User' in the proposed ontology. 

We design the \textit{User Profile} to cover the interest, intention and behavioural aspects defined in \cite{GaoLW10}. Those are represented by the classes `Learning Preference', `Learning Goal', `Academic Parameter', and `Psychological Parameter'. The `Academic Parameter' captures the learner's performance, such as test scores, while the `Psychological Parameter' reflects the state-of-mind of the learner, such as being tired. This focus on the psychological state is due to its influence on the overall learning process and performance. 








\section{Use Case Scenario}\label{usecase}
We describe a general and a specific use case. In a general use case, an OER repository owner could utilise the EduCOR ontology to model the learning materials in their repository. 
The repository serves learners through a standard search and information retrieval functionality. In the future, it could be possible to integrate an automatic decision-support system with minimum to zero adjustments of the repository structure.

We also used our ontology in specific use case, in the development of eDoer\footnote{\url{http://edoer.eu}\label{eDoer}} platform, an open learning recommender system prototype, focusing on Data Science related jobs \cite{tavakoli2020oer,tavakoli2020recommender,tavakoli2020labour}. Since eDoer aims to empower learners through open, personalised learning and curriculum recommendations based on labour market information and OERs, the following components have been deployed using the EduCOR ontology: 1) we used the \emph{Skill} pattern to bridge between jobs and their required qualities, 2) we applied the \emph{Knowledge Topic} pattern 
to decompose each skill into relevant learning components, 3) the \emph{Learning Path} pattern 
was used to create a path for learners which includes a sequence of knowledge topics towards their learning goals (i.e. target job or skills), 4) to store the required learning resources into our system, we applied the \emph{Educational Resource} pattern
, 5) in the process of building a personalised learning content recommender engine, we benefited from \emph{Recommendation} and \emph{User Profile} patterns to offer the most relevant learning items (i.e. knowledge topics and learning materials) to learners based on their learning goals, learning preferences, and their current knowledge level, and 6) the \emph{Test} pattern was used to offer assessment services in order to help learners to monitor their progress towards their learning goals. 

Therefore, on the eDoer platform, learners can set their target job, and the system will provide them with a list of skills they need to master for that particular job. Learners are offered to select one or more of those skills and set them as learning objectives. Moreover, learners can search through other existing skills and add additional learning goals. They can also set their learning preferences, such as type of learning materials, and the length of content, which results in personalised learning content recommendations. The generated learning path includes the target skills and the necessary knowledge topics covered for each skill. Subsequently, users receive OERs for each of the knowledge topics, which can be viewed, rated, and changed. Based on the users' feedback (i.e. ratings) on each of the recommendations, eDoer updates the users' preferences to capture any changes in user preferences. Moreover, there are various assessments available both on skill and knowledge topic levels that provide means to monitor the learning process\footnote{You can watch a demo of eDoer here: \url{https://youtu.be/5PRcUgNa7tA}}.
Up to now, we evaluated eDoer in the context of a \emph{Business Analytics} course at the \emph{University of Amsterdam}. This evaluation revealed that 24 students out of 97, who worked with our system voluntarily, achieved higher course grades than those that did not.

\section{Evaluation} \label{evaluation}

Several evaluation methods have been introduced in the literature on ontology development. A recent survey~\cite{Ivanova2020Ontology} classified evaluation methods under five main approaches: 1) Gold-standard based, 2) Corpus-based or data-driven, 3) Task-based of Application-based, 4) Criteria-based, and 5) Evaluation by humans. 

To ensure objectivity when evaluating EduCOR, we decided to use inductive methods following \cite{chari2020explanation,li2010ontology} to select the most relevant evaluation criteria for our proposed ontology. Therefore, based on \cite{brewster2004data,degbelo2017snapshot}, we focus on coverage and adaptability as key performance indicators (KPIs) of the EduCOR ontology. In the context of ER representation for learning-material repositories, the coverage is defined as the ability to describe learning materials by classes. Adaptability is defined as the potential to represent multiple repositories homogeneously. 
Based on these two KPIs, we conduct the gold-standard and task-based evaluation approaches. The gold-standard valuation is meant to compare EduCOR directly to other repository schemata, while the task-based evaluation is meant to validate its performance in real-world use cases. We also evaluated the proposed ontology design with experts in the ontology development domain.

\subsection{Gold standard-based evaluation}
To measure EduCOR's coverage and adaptability towards other existing ontologies, we select three well-established repositories for ER resources, namely Merlot\footnote{\url{https://www.merlot.org}}, SkilsCommons\footnote{\url{https://www.skillscommons.org}}, and OERCommons\footnote{\url{https://www.oercommons.org}}. Since those repositories' APIs are not open, we conduct a thorough analysis of repositories' schemas based on the information on their websites, user guides, and the use of hosted materials and resources. We extract the overall class representations of the three schemas. Ultimately, these schemas are accepted as gold standards, against which the EduCOR ontology is compared. The comparison is conducted by mapping EduCOR classes to the underlying schema of each repository. Mapping refers to identifying classes in the gold standards that classes in EduCOR can represent. Since mapping is dependent on the clear definition of a schema's own vocabulary, it may lead to a subjective evaluation. Therefore, we conduct the mapping as a multi-fold process, in which four different developers assess the meaning of the classes in the proposed ontology and the compared schema. Once the mapping process is conducted, we seek a tangible representation of the coverage and adaptability metrics. To accomplish this task, we follow the work of \cite{brewster2004data} to calculate the recall based on the definition from the information-retrieval domain to represent the coverage of EduCOR. In this adaption, we define the true positive value as the number of classes covered by EduCOR and existing in the gold schema, while the false negative is defined by the number of classes in the schema that EduCOR did not cover.
The calculated recall values are given in Table~\ref{tab:recall_values}. They indicate the ability of the EduCOR ontology to represent data in the selected repositories with a high coverage level of more than 83\%.

\begin{table}
\vspace{-3.5mm}
\centering
\caption{Recall values of EduCOR as calculated for each gold schema}
\label{tab:recall_values}
\setlength{\tabcolsep}{8pt}
\begin{tabular}{|P{3cm}|P{2.7cm}|P{2.5cm}|P{1.7cm}|}
\hline
&  {\bfseries OER-Commons} & {\bfseries SkillsCommons} & {\bfseries Merlot}\\
\hline
{\bfseries EduCOR ontology} & 0.833 & 0.857 & 0.875 \\
\hline
\end{tabular}
\vspace{-3.5mm}
\end{table}

To evaluate adaptability, we refer to the definition as mentioned earlier of this measure in the context of ER repositories. Here we qualitatively assess the ability of EduCOR to represent three different repositories, which have distinct differences in focus when representing the ERs and OERs. Examples of those differences include the emphasis of Merlot on user roles, the links in SkillsCommons between ERs and industrial occupations, and the focus on educational and evaluation standards in OER-Commons. Despite those differences, our proposed ontology homogeneously represented them all, with high recall values. Moreover, EduCOR ontology provides other repositories with additional features in terms of learning material representation, user modelling and learning recommendations. This can be seen from linking ERs and OERs to the labour market through the `Skill' class, the inclusion of `Psychological Parameter' in the user profile; and through the `Recommendation' and `Learning Path' classes that enable a personalized learning experience. 

\subsection{Task-based evaluation}
In this step of the ontology evaluation, we defined specific tasks and evaluated EduCOR's ability to fulfil them. For the task-based evaluation, we followed the approach of \citet{chari2020explanation}, where competency questions are defined to reflect the main contributions of EduCOR, based on a sample use case that is expected to be executed by a potential user of the ontology. Such a use case is described as a general use case in Section~\ref{usecase}. It is one of the contributions of EduCOR in representing ERs and OERs from multiple sources and enabling user-centric, job market-oriented learning recommendations. 
From the previous use case, 
we define three main tasks that EduCOR should fulfill:
\begin{enumerate}
\item Adaptable representation of OERs from multiple sources.
\item Consideration of labour market skills in the learning path.
\item User-centred design, considering learner's academic and psychological needs within the user profile.
\end{enumerate}

To evaluate EduCOR's ability of performing these tasks, the following set of questions were designed: 
\begin{itemize}
    \item Q1: How to retrieve OERs from multiple sources for a learning goal?
    \item Q2: How can a personalized OER difficulty be chosen for the user?
    \item Q3: How to provide an OER to a user with a specific access mode?
    \item Q4: How to retrieve required OERs for a certain job skill?
    \item Q5: What is required to generate a personalized learning path?
    \item Q6: How to personalize a learning recommendation based on a user's psychological state?
\end{itemize}
The first question Q1 reflects the adaptability metric in the evaluation of the ontology. Questions Q2 and Q3 focus on the personalisation of the retrieved material towards specific user needs, such as the difficulty levels and accessibility modes. Those questions represent the richness in data-type properties, which scaffolds the personalisation of retrieved or recommended ERs and OERs. Q4 evaluates links that the ontology draws between the ERs and the labour market needs. 
This allow the ER repository developer to support the users wit career-oriented recommendation.
Q5 and Q6 evaluate the user-centricity of the ontology in terms of representing the user's academic and psychological parameters in a recommendation or the retrieval of ERs. 
These parameters are important as they reflect the user's status, mentally and academically, which allows the recommendations to be more tailored towards their actual needs from the ERs.
These competency questions are directed to the EduCOR ontology through SPARQL queries, where their answers are retrieved from the available data. A sample SPARQL query is provided in Listing 1.1. The full description of queries and their answers are accessible on the documentation web page.



\begin{lstlisting}[basicstyle=\footnotesize\ttfamily,frame= lines, caption=SPARQL query to answer the competency question Q2]
PREFIX ec: <https://github.com/tibonto/educor#>
PREFIX dc: <http://purl.org/dcx/lrmi-vocabs/alignmentType/>

SELECT * 
WHERE {
    ?test           ec:testKnowledgeTopic   ?knowResource.
    ?knowResource   ec:difficulty           ?difficulty.
    ?user           ec:solves               ?test.
    ?user           ec:hasProfile           ?userProfile.
    ?acadParam      ec:storedIn             ?userProfile.
    ?acadParam      dc:educationalLevel     ?currentLevel. 
}
\end{lstlisting}

\section{Related Work}\label{relatedwork}

Ontology development for the educational domain is not a new task. Many ontologies have been developed in the last years related to education systems and learning materials~\cite{TarusNM18}. However, we find a series of issues that dated published ontologies have, such as maintainability, online availability, metadata, and their quality\footnote{An example is the Medical Educational Resource Aggregator \url{https://bioportal.bioontology.org/ontologies/MERA}}. The biggest challenge is that most of the relevant works are not publicly available anymore. Another critical factor to consider is that the main interest in educational domain ontologies comes from educators and non-technical personnel. Therefore, the majority of these ontologies focus on educational perspectives rather than rich metadata.

In the plethora of educational and e-learning ontologies, we find the majority of ontologies in the domain of application, or task-specific. Only a small minority were developed to describe the learning domain and learner data~\cite{StancinPJ20}. This creates a challenge in adopting such ontologies to general settings and applications. Such an example could be the recent work in ontology-based curriculum mapping by \citet{ZouriF21}, which is focused on creating a core ontology for curricula and courses in higher education institutions. Such an ontology raises important challenges when trying to  fit in a general purpose e-learning environment as they cannot be accurated mapped to another domain. General domain educational ontologies are closely related to our goal; hence, we focus our analysis there.

 \citet{koutsomitropoulos2018learning} create an ontology-based on the IEEE LOM standard and SKOS for OER repositories. They propose an enhancement of the ER's metadata, and they link to thesauri dataset. However, they offer no personalised content capabilities. Recently \citet{ChimalakondaN20} suggested ``an ontology based modelling framework for design of educational technologies''. Similar to their model, we include context and domain-specific ontology to our design, and add the ``GoalsOntology'' as `Learning Goal' in our system. However, in contrast to their framework, our design offers personalised recommendation features.

Another related domain in the literature are 
personalised recommendation systems.
\citet{bulathwela2020truelearn} propose an OER recommendation system based on learner background knowledge and content but without an underlying ontology. However, recent reviews show the growing significance of personalisation and recommendation systems in e-learning models, and ontologies are proven to be useful in this respect \cite{GeorgeL19}. \citet{jando2017personalized} show that most techniques use such an ontology to accomplish personalisation, such as the work in~\cite{HarrathiTB17, jeevamol2021ontology}. A review by \citet{TarusNM18} presents the state-of-the-art for ``ontology-based recommenders in e-learning''. It points out the gained popularity of e-learning resource-recommendations and ``their ability to personalise learner profiles based on the learner's characteristics, such as background knowledge, learning style, learning paths and knowledge level''. It is noticeable from the state-of-the-art that despite the variety of ontology-based recommender systems in the last years, only the most recent works have developed the ontology in OWL or RDF and offer metadata description. Moreover, the vast majority of publications use an ontology as a tool that provides information to a recommendation algorithm rather than integrating recommendation requirements in the ontology itself. 
We address this issue in EduCOR by integrating recommendation class with the overall representation of ERs and user profile.

In terms of connecting the labour market representation with an educational ontology, one of the most related approaches is the ``Ontology-based personalised course recommendation framework'' by \citet{ibrahim2018ontology}, which uses a course, a student and a job ontology to recommend courses and jobs. Inspired by their design, we divided the student ontology into user profile and skill pattern, offering personalisation capabilities, such as the `Learning Preference' class.

\begin{table}[!h]
\centering
\caption{Table comparison of the related work compared to EduCOR}
\label{tab:comparison_table}
\begin{tabular}{| m{2.6em} | m{0.8cm} |m{1cm}| m{1cm} |m{5.4cm}| m{2.4cm} |}
\hline
Paper & FAIR & Evalu- ation  & Data availability & Personalisation & Reuse of vocabularies    \\
\hline
\cite{ChimalakondaN20} & No & Yes & Yes & Goals (Learning goals) & No   \\ [0.2cm]

\cite{HarrathiTB17} & No  & No & No  & Learning preferences, Learning style, Learner characteristics, Knowledge level, Learning activities & W3C recommendation ontology  \\ [0.2cm]

\cite{ibrahim2018ontology}  & No  & Yes & No  & Education information, Job related skills  & No  \\ [0.3cm]

\cite{jeevamol2021ontology} & No & No & No & Learning Style, Learning pathways  & IEEE LOM \\ [0.2cm]

\cite{koutsomitropoulos2018learning} & No & Yes & No & Datatype properties & IEEE LOM, thesauri, SKOS vocabulary \\ [0.2cm]

\cite{skillen2014ontological} & No & Yes & No &  Accessibility, Activities, Health conditions & No   \\ [0.3cm]

\cite{ZouriF21} & No & No & No & Learning pathways & No      
\\ [0.2cm]
\hline
Ours & Yes & Yes & Yes & Learning Goal, Learning pathways, Accessibility, Learning preferences, Psychological parameter, Academic parameter, Recommendation, Datatype properties & IEEE LOM, CCSO, DCMI, SKOS, schema.org \\
\hline
\end{tabular}
\end{table}

User modelling plays an essential role in ontology-based recommendations~\cite{GeorgeL19} since the information about the user is vital to personalise the recommendation itself. \citet{EkeNSN19} present a comprehensive review on user modelling and argue that ontologies are the best solution to unify the user profile representation. 
\citet{GaoLW10} categorise user modelling approaches under three main classes: behavioral modelling, interest modelling and intention modelling. They show that personalisation is based on these three pillars. User profiling and content modelling are both considered inputs to a filtering algorithm, such as a recommendation system, to generate a personalised output.
The content of user profiles has also been witnessing increased attention in recent years. This is also influenced by the ability to transfer the user profiles among multiple applications and domains~\cite{EkeNSN19}. In the educational domain, not only the academic parameters are essential in generating personalised recommendations, but also the psychological parameters, as pointed out by \citet{Fatahi19}. This importance is shown in their adaptive e-learning environment study, where they showed enhanced student performance when receiving personalised recommendations. Students in their study also showed mire attraction to the personalised system, since it ``can understand their emotional state better''. 
Further, the authors in \citet{skillen2014ontological} developed an ontological representation of users, putting a focus on their psychological health conditions alongside their learning-related preferences and activities. We found these previous approaches necessary in the educational field. Therefore, we expanded and complemented this set of ontological user profiling works by proposing a hybrid representation in EduCOR. As a result, in our user profile pattern, static and dynamic parameters represent both academic and psychological aspects of the learner. 

Table~\ref{tab:comparison_table} shows a summary of the comparison between EduCOR and the above-mentioned related educational ontologies. From this summary, one can notice that EduCOR exceeds the state of the art. It is aligned with FAIR data principles, and provides richer perosonalization features, both in classes and datatype properties, compared to the related work ontologies. Furthermore, the EduCOR extends these works by embedding the `Recommendation' and `Skill' classes in a unified representation, offering stronger links between the ERs and personalised recommendations. 

\section{Discussion and Future Steps}

EduCOR is a publicly available, findable, registered\footnote{You can find EduCOR's presentation at \url{http://ontology.tib.eu/educor} and on our GitHub page at \url{https://github.com/tibonto/educor}}, and lightweight ontology that can host ERs and OERs, personalised recommendation system features, and user profiles. It is created to address the gap between the educational domain, the labour market, and personalised learning. EduCOR can be used as a whole or as parts via the patterns introduced in Section~\ref{coreonto}. It is a semantically enhanced ontology that is adaptable. Therefore, EduCOR can be used in different educational domains, such as Computer Science, to support online learning platforms and personalised education systems. 
 EduCOR is enriched with the necessary vocabulary and rich metadata to be general enough to be used in different settings. We leverage and maintain compatibility with existing educational repositories related to Massive Open Online Courses (MOOC) and OERs, as shown in Section~\ref{evaluation}. Moreover, we expand on them to include personalised representational primitives needed for modelling the components of a recommendation system.
 
However, EduCOR does not provide data specific 
to an application domain, and expert intervention may be necessary to seamlessly align the domain-specific ontology to the EduCOR ontology. Also, EduCOR does not offer automatic mapping of courses and curricula to its ontology. Although, this can happen by identifying courses, or chapters' learning objectives, and classifying them in skill categories with corresponding knowledge topics. An automatic alignment system for domain and task-specific ontologies mapping to EduCOR ontology is also part of future efforts.


We have implemented the basic ER and OER components that are necessary to link with the labour market and offer personalised learning. However, some aspects of OERs, and the recommendation system might need more thorough analysis. We foresee EduCOR extensions to include further analysis of some classes. The quality indicators could extend to summarize the resource multimedia and metadata quality with user's feedback ratings. Another extension could be the analysis of learning preferences, which could further link to special education coverage. Also, the accessibility analysis could expand to offer additional representations in our system, by covering user accessibility, preferences, and content access rights. This work will additionally aim to assist in the user privacy and profile restrictions alignment \hbox{with our ontology.}  


In future work, we plan to publish an Open Educational Knowledge Graph, connecting educational resources with the labour market while offering personalised recommendation features by combining ERs from multiple sources. Upon identifying the appropriate content and repositories, we wish to gather the requirements and publish the Knowledge Graph based on the EduCOR ontology.
Therefore, we foresee a sustainability plan for the following years as we plan to use the EduCOR ontology as the basis of our future work. We are committed to its maintenance and extensibility to address future challenges and meet future requirements.

\section{Conclusion}

We have built an open-source, free access ontology aimed at modelling the educational domain, personalised learning recommendation systems, user profile, and labour market data. We argued that this interdisciplinary attempt is vital both for the SW, educators and the broader community.
Our requirement analysis came from reviewing the literature and an existing e-learning system that revealed the key components of a perspective system around which we built our ontology. We presented our design and ontological components, which adopt open community standards and FAIR data principles. 
We evaluated EduCOR with gold-standard and task-based approaches and showed that the EduCOR ontology achieves high coverage to multiple OER repositories. Through a carefully crafted set of competency questions, we evaluated the capabilities of EduCOR in assisting the system designers in e-learning based recommendation systems to determine the necessary elements for their system. 
We believe our ontology can be a beneficial tool for system designers as they implement personalised features in their recommendation system. We are committed to continuing this line of work towards supporting future requirements that would extend our ontology.

\section*{Acknowledgements}
  We would like to thank Dr. Javad Chamanara for setting up the ontology page. This work was partially funded by the European Union's Horizon 2020 research and innovation program under the Marie Sk\l{}odowska-Curie project Knowgraphs (grant agreement ID: \href{https://cordis.europa.eu/project/id/860801}{860801}), the European Research Council for the project ScienceGRAPH (grant agreement ID: \href{https://cordis.europa.eu/project/id/819536}{819536}) and the ERASMUS+ Key Action 204 Higher Education project OSCAR (grant agreement ID: \href{http://oscar-ai.eu/}{2020-1-DE01-KA203-005713}).

\bibliographystyle{splncsnat}
\bibliography{bibliography}

\end{document}